\documentclass[sigconf,nonacm,natbib=false]{acmart}
\AtBeginDocument{%
  }

\setcopyright{rightsretained}
\renewcommand\footnotetextcopyrightpermission[1]{}

\usepackage{siunitx}
\usepackage{enumitem}
\usepackage{float}
\RequirePackage[
  datamodel=acmdatamodel,
  style=acmnumeric,
  ]{biblatex}

\begin{document}
\title[An Analysis of Ethereum Builder Centralization]{Order Flow Exclusivity and Value Extraction Mechanisms: \\An Analysis of Ethereum Builder Centralization}

\author{Ao Zhang}
\affiliation{%
  \institution{Tsinghua University}
  \city{Beijing}
  \country{China}}
\email{zhang-a20@mails.tsinghua.edu.cn}

\author{Yunwen Liu}
\affiliation{%
  \institution{KU Leuven}
  \city{Leuven}
  \country{Belgium}}
\email{yunwen.liu@esat.kuleuven.be}

\author{Ren Zhang}
\affiliation{%
  \institution{Cryptape and Nervos}
  \country{China}}
\email{ren@cryptape.com}

\author{Yingdi Shan}
\affiliation{%
  \institution{Tsinghua University}
  \city{Beijing}
  \country{China}}
\email{shanyd@tsinghua.edu.cn}

\author{Yongwei Wu}
\affiliation{%
  \institution{Tsinghua University}
  \city{Beijing}
  \country{China}}
\email{wuyw@tsinghua.edu.cn}

\renewcommand{\shortauthors}{Zhang et al.}

\begin{abstract}
This study investigates the rapid centralization of the Ethereum builder market under the Proposer-Builder Separation (PBS) architecture.
We argue that existing research, by focusing predominantly on influential order flows, lacks a comprehensive evaluation of order flow behavioral patterns and economic purposes.
To address this gap, we analyze Ethereum transactions from September 2023 to August 2025 to characterize Exclusive Order Flows (EOFs) and non-atomic Maximal Extractable Value (MEV)---the missing components corresponding to these behavioral and economic dimensions, respectively.
We introduce a novel exclusivity metric based on Kullback-Leibler divergence and employ supervised learning to identify 75 EOFs and 322 non-atomic MEV flows, which account for 71\% and 23\% of trading-related builder revenue.
A longitudinal analysis of builder strategies across these dimensions delineates the market's evolution into four distinct eras, revealing that while EOFs were instrumental in establishing early dominance, incumbents have since decoupled market share from immediate EOF dependency by leveraging entrenched network effects.
Ultimately, we conclude that builder centralization is an emergent property of the PBS framework itself, as the architecture systematically violates the fundamental prerequisites of a competitive market.

\end{abstract}

\maketitle

\section{Introduction}
\label{sec:intro}
Since Ethereum's transition to proof-of-stake in September 2022~\cite{grandjean2024ethereum}, \emph{proposer-builder separation (PBS)}~\cite{heimbach2023ethereum} has emerged as the dominant block production architecture, accounting for over 92.5\% of all blocks as of December 2025~\cite{mevboost}.
Under this paradigm, consensus participants, or \emph{proposers}, no longer collect transactions or construct blocks directly.
Instead, specialized agents known as \emph{searchers} collect transactions from users and bundle them to extract value, subsequently passing these bundles to \emph{builders}.
Builders then aggregate these bundles into block candidates and compete for inclusion by offering the highest \emph{bid} to the proposer.
This bid represents a portion of the block-construction revenue, with the builder retaining any remaining surplus as \emph{profit}.
Given that Ethereum transactions frequently interact with smart contracts, a stream of transactions targeting a specific contract is called an \emph{order flow}.


The Ethereum builder market has rapidly transitioned from a distributed landscape into a rigid, centralized structure.
In late 2022, no single builder controlled more than 20\% of the market share~\cite{bahrani2024centralization}; however, by September 2024, a small oligopoly dominated the network, with Titan and Beaverbuild capturing a combined 87.7\%~\cite{mevboost}.
By December 2025, while the emergence of BuilderNet absorbed Beaverbuild's share, the underlying market concentration remains entrenched. 
Such centralization hinders technical progress as the key to builder success has migrated from technical innovation to the formation of exclusive coalitions with top-tier searchers.
Moreover, it compromises the network's security by introducing single points of failure, undermining the core principle of decentralization, and threatening Ethereum's censorship resistance.
Consequently, numerous studies have sought to analyze these market dynamics to understand the forces driving this concerning evolution.


However, existing research predominantly focuses on \emph{influential order flows} that contribute the largest proportion of revenue to the leading builders~\cite{gupta2023centralizing,heimbach2024non,oz2024wins, yang2025decentralization}.
We argue that this approach cannot fully capture the market dynamics leading to builder centralization.
Specifically, analyzing only influential order flows offers limited insights into two critical dimensions:

\smallskip\noindent\textbf{Exclusivity.} Certain order flows are exclusively accessible to a limited subset of builders~\cite{oz2024wins}, or in some cases, a single builder~\cite{heimbach2024non}.
These \emph{Exclusive Order Flows (EOFs)} are deemed a primary driver of builder centralization, manifesting in a ``chicken-and-egg'' dilemma: since public order flows are accessible to all competitors, builders require EOFs to generate profit, yet they only attract such flows after securing a significant market share~\cite{oz2024wins}.
This structural barrier prevents new entrants from gaining traction. 

However, current analyses of EOFs suffer from two limitations.
First, because existing studies focus exclusively on the leading builders and their respective EOFs, the extent to which smaller or emerging builders possess their own EOFs remains unknown.
This lack of granular data obscures the overall contribution of EOFs to the builder market.
Second, the lack of temporal analysis prevents an exploration of how dominant builders initially acquired their market share, offering limited insight into the causal relationship between EOFs and the evolution of market dominance.


\smallskip\noindent\textbf{Value Extraction Mechanisms.} Drawing upon recent studies~\cite{gupta2023centralizing, heimbach2024non, oz2024wins, yang2025decentralization}, we categorize order flows into three value extraction mechanisms based on their economic intent:
(1)~\emph{protocol} encompasses a diverse array of decentralized finance applications;
(2)~\emph{atomic MEV} involves strategies where the initiator captures immediate profit within a single block through circular trades or liquidations~\cite{oz2024wins,qin2022quantifying};
(3)~\emph{non-atomic MEV} consists of non-circular token swaps where profits are realized across different venues, such as arbitrage between decentralized and centralized exchanges~\cite{gupta2023centralizing,heimbach2024non}. 

While the former two categories can be identified through automated methods or public labels, non-atomic MEV remains under-documented; only thirteen instances have been identified~\cite{heimbach2024non}.
The absence of a scalable identification mechanism for non-atomic MEV precludes a rigorous analysis of the value extraction strategies employed by leading builders.
Consequently, it remains unclear whether or how their strategies differ from those of the others.

\smallskip
We observe that the limitations of these studies are rooted in their methodology: they identify influential order flows by isolating transactions with distinct features~\cite{gupta2023centralizing, heimbach2024non} or utilizing auxiliary information~\cite{oz2024wins, yang2025decentralization}, such as known Telegram bots.
While these heuristics can identify specific EOFs or non-atomic MEV instances, they cannot guarantee the comprehensiveness required for a systemic analysis.
These methodological gaps call for a comprehensive examination of \emph{all} order flows, categorized by their exclusivity and value extraction mechanisms, to fully understand the builder market landscape and its evolution.

In this paper, we address this need by introducing the missing components for identifying both EOFs and non-atomic MEV order flows.
These identification mechanisms enable a comprehensive analysis of the two aforementioned dimensions, which together provide a holistic perspective on order flow dynamics.
Specifically, an order flow can be characterized by its \emph{behavioral patterns} and its \emph{economic purpose}, corresponding to exclusivity and value extraction mechanisms, respectively.
This integrated framework allows us to rigorously explore the underlying dynamics contributing to builder centralization.
Our contributions are as follows:

\smallskip\noindent\textbf{Characterizing Exclusive Order Flows.}
We conduct an exhaustive scan of all Ethereum transactions between September 1, 2023, and August 31, 2025, identifying 164\ 249 order flows.
To evaluate these flows, we define a novel \emph{exclusivity} metric designed to mitigate the two primary statistical biases: \emph{endogeneity bias}, where high market share leads to disproportionately high order flow capture, and the \emph{sample size confound}, which introduces stochastic errors in low-frequency transactions.
Specifically, our metric incorporates the Kullback-Leibler (KL) divergence between an order flow's bribe distribution among builders and the aggregate builder block-construction market distribution to counteract endogeneity bias.
We further weight this divergence by the flow's total generated bribe to eliminate the sample size confound.

Applying this metric, we identify 75 EOFs that account for 70.53\% of total trading-related revenue.
Notably, 68 of these were previously unreported, contributing 34.96\% of the total EOF bribe.
The efficacy of our automated approach is underscored by the identification of 37 EOFs active between September 2023 and March 2024, a period extensively covered by prior studies~\cite{heimbach2024non, oz2024wins, yang2025decentralization} that failed to detect them.
Our analysis further reveals a strategic heterogeneity: while market leaders maintain diverse income streams and do not rely solely on EOFs for profitability, the long tail is sharply divided between small builders who are entirely dependent on exclusive flows and those with no access to them at all.

\smallskip\noindent\textbf{Identifying Non-Atomic MEV Order Flows.}
To automate the discovery of non-atomic MEV, we develop a supervised learning framework employing a decision tree classifier.
We manually labeled the value extraction mechanisms of the top 210 revenue-contributing order flows to establish a ground-truth dataset.
Our model utilizes an expanded feature set that augments standard transaction-level data with aggregate statistical features of the entry-point contract, such as the count of unique interacting addresses and the average transaction frequency per sender address.
This approach achieves an identification accuracy of 92.06\%.

Our methodology uncovers 322 non-atomic MEV order flows, of which 316 are previously undocumented.
This comprehensive mapping allows us to quantify the economic significance of non-atomic MEV, which collectively accounts for 22.99\% of total builder revenue; notably, our newly identified flows contribute 39.84\% of this subtotal.
We observe a stark winner-take-most dynamic within this category, as the top two order flows alone capture 56\% of all non-atomic bribes.
Despite this concentration, the bribes follow a heavy-tailed distribution with a power-law exponent of $\alpha = 1.47$.
This slow decay indicates that a persistent long tail of niche players remains active.
Furthermore, an analysis of the trading pairs utilized across these flows reveals a diverse spectrum of behavioral complexity across both high-volume and niche order flows.


\smallskip\noindent\textbf{Analyzing the Temporal Evolution of Builder Centralization.} Leveraging the granular data from our multi-dimensional classification, we analyze the longitudinal evolution of the builder market.
By calculating the weekly Herfindahl-Hirschman Index (HHI) of builder market shares, we observe a distinct trajectory toward centralization.
We delineate this evolution into four eras: \emph{Genesis}, \emph{Algorithm Wars}, \emph{EOF Moats}, and the contemporary \emph{Oligopoly}, defined by shifts in HHI and the prevailing competitive advantages of market leaders.
Our analysis of the latter three eras reveals that while atomic and non-atomic MEV remain critical revenue drivers, non-atomic MEV exhibits consistent growth across all periods.

To investigate the causal dynamics of the ``chicken-and-egg'' dilemma, we evaluate the Pearson correlation between builders' market shares and their respective EOF bribes.
Our findings reveal shifting dependencies: dominant builders such as Titan and Beaverbuild exhibited strong correlations with EOF bribes during the \emph{EOF Moats} era, yet this relationship decoupled during the \emph{Oligopoly} era.
This suggests that while EOFs were instrumental in establishing their initial dominance, these leaders no longer require steady EOF income to sustain their market positions, effectively solidifying their status as entrenched incumbents.


\smallskip
In summary, our analysis demonstrates that EOF and non-atomic MEV are \emph{symptoms} and \emph{catalyst} of a broader oligopolistic trend rather than its root cause.
These mechanisms are not the exclusive domain of leading builders, nor does a revenue increase in these areas provide a deterministic guarantee of market share growth.
Instead, we conclude that builder centralization is an emergent property of the PBS architecture itself.
By violating three fundamental prerequisites for a competitive market---diminishing returns to scale, information symmetry, and low entry barriers---the PBS framework facilitates an inevitable trajectory toward centralization.
\section{Ethereum’s Builder Centralization Problem}
\label{sec:background}

This section provides the necessary background on Ethereum transaction confirmation and the Proposer-Builder Separation (PBS) architecture (Sec.~\ref{subsec:pbs}).
We then classify the primary mechanisms for value extraction (Sec.~\ref{subsec:value_extraction}).
Finally, we evaluate the existing literature to pinpoint the analytical gaps that motivate our research questions (Sec.~\ref{subsec:gaps}).


\subsection{Proposer-Builder Separation Framework}
\label{subsec:pbs}
Ethereum utilizes two types of accounts: \emph{Externally Owned Accounts (EOA)}, controlled by private keys, and \emph{Contract Accounts (CA)}, governed by their executable code.
Crucially, only an EOA can initiate a state transition by signing a \emph{transaction}.
Such transactions may involve a simple transfer of ETH---the protocol’s native currency---or target a specific function within a destination CA.
In the latter case, the CA can trigger a sequence of nested calls to other contracts.

A primary application of this architecture is the \emph{Decentralized Exchange (DEX)}, a protocol implemented via CAs that facilitates permissionless asset exchange.
The Ethereum ecosystem hosts multiple DEXes---e.g., Uniswap, Curve, and Balancer---each comprising numerous \emph{pools}.
Each pool is responsible for a specific trading pair, and every individual \emph{swap}---the trading of one token for another---targets a specific pool to execute the exchange.
These swaps occur either through direct interaction with a DEX pool or via an aggregator CA that orchestrates multiple nested calls across various venues to optimize execution.
From the perspective of block production, these activities generate two distinct revenue streams: the \emph{priority tip}, paid via the protocol's native transaction fee mechanism, and the \emph{direct bribe}, transferred to the builder directly or through smart contract execution.
This direct bribe typically originates from the \emph{Maximal Extractable Value (MEV)}---the total value realized through the strategic ordering or inclusion of transactions.
For brevity, we use \emph{transaction bribe} to denote the aggregate value of the priority tip and direct bribe triggered by a transaction.

The PBS framework was introduced to mitigate the centralization risks associated with MEV by decoupling the role of transaction sequencing from block proposal~\cite{grandjean2024ethereum,heimbach2023ethereum}. 
While natively envisioned for the consensus layer, PBS is currently realized through the \emph{MEV-Boost} middleware, which, as of December 2025, facilitates the production of over 92.5\% of blocks~\cite{mevboost}.
The ecosystem defined by this out-of-protocol implementation involves five key participants:

\begin{description}
\item[Users] initiate state transitions by submitting transactions. While these are typically broadcast to the public peer-to-peer network, i.e., the mempool, transactions may also be sent to builders via private channels.
\item[Searchers] extract value by identifying profit opportunities, such as arbitrage, within the stream of pending transactions. They aggregate these opportunities into \emph{bundles}---an ordered sequence of transactions---and submit them to some or all builders via private APIs.
\item[Builders] function as the primary architects of the block. They collect bundles from searchers and transactions from the mempool to construct full block bodies, optimizing for total revenue. Builders then compete for the right to produce the next block by submitting bids to proposers via relays.
\item[Relays] act as mutually trusted intermediaries between builders and proposers. They verify the validity of the blocks submitted by builders and forward only the highest-bid block headers to the proposers. Upon receiving a signed header from the proposer, the relay reveals the full block body to the proposer and assists in propagating the block to the network.
\item[Proposers] are the consensus participants responsible for the final inclusion of a block. Under PBS, they no longer need to perform complex transaction ordering; instead, they simply sign the most profitable block header provided by a relay to earn the associated bid.
\end{description}

Throughout this paper, we define \emph{bribe} as the aggregate transaction bribe generated by a specific order flow (defined in Sec.~\ref{subsec:bribe}).
A builder's \emph{revenue} represents its total gross income from all bribes, whereas its \emph{profit} in a block is the revenue remaining after subtracting the bid paid to the proposer.
Notably, a builder's profit may be negative when the bid exceeds the revenue.

\subsection{Value Extraction Mechanisms}
\label{subsec:value_extraction}
Builder revenue is primarily derived from trading-related activities.
Following existing literature~\cite{gupta2023centralizing, heimbach2024non, oz2024wins, yang2025decentralization}, we categorize these extraction mechanisms based on the destination CA of the revenue-generating trades:

\begin{description}
\item[Protocol.] This category encompasses standard user interactions with decentralized finance (DeFi) applications, such as token swaps or lending protocol operations.
While the majority of this activity is public, specialized protocols can generate private order flows.
For instance, \emph{intent-based protocols} like CoWSwap or 1inch Fusion allow users to submit high-level declarative demands rather than specific transaction steps.
These protocols rely on a network of third-party agents to satisfy these intents by generating the necessary swaps, resulting in order flows that are often sent exclusively to preferred builders.

\item[Atomic MEV.] Atomic MEV refers to risk-free extraction strategies where an entire sequence of trades is executed within the same block. Common examples include atomic arbitrage~\cite{qin2022quantifying, wang2022cyclic}, which balances prices across on-chain DEXes, and sandwich attacks, which involve placing trades immediately before and after a targeted user transaction.
These strategies are transparent because the profit is realized entirely on-chain.
Such deterministic token-flow patterns make atomic MEV identifiable through static or trace analysis.

\item[Non-Atomic MEV.] Non-atomic MEV, primarily represented by CEX-DEX arbitrage or cross-chain DEX arbitrage, involves executing one leg of a trade on-chain while the offsetting leg occurs in an external environment, such as a centralized exchange or another blockchain~\cite{heimbach2024non}.
Unlike atomic strategies, non-atomic MEV transactions appear on the Ethereum ledger as simple, non-circular swaps because the balancing trade is invisible to the public.
These strategies carry inventory risk and require proprietary capital, making them mimic ordinary user behavior. Consequently, identifying non-atomic MEV is challenging.



\end{description}

\subsection{Existing Studies and Analytical Gaps}
\label{subsec:gaps}
While the PBS framework successfully democratized MEV rewards for proposers, it fundamentally shifted centralization pressure to the builder layer.
Specifically, builders with superior hardware, optimized algorithms, or privileged access to private order flows often consistently outbid their peers.
This competitive landscape has resulted in a highly concentrated market where an entrenched oligopoly frequently controls a majority of the block production.
Existing empirical research has largely focused on analyzing dominant builders by tracking influential order flows from known entities or high-volume sources.
Next, we review the key findings relevant to our study and identify the unaddressed research questions.


\smallskip\noindent\textbf{Gap 1: The Granularity-Coverage Dilemma in Order Flow Definition.}
Existing definitions of an order flow typically struggle to reconcile the need for granular insight with the requirement for comprehensive coverage.
One prominent line of research~\cite{bahrani2024centralization, gupta2023centralizing, wang2024private, yang2025decentralization} distinguishes only between public and private order flows based on visibility.
While useful for high-level analysis, this binary classification is too coarse to discern individual searcher-builder strategies or specific value extraction mechanisms.
Conversely, defining order flows by individual sender addresses~\cite{heimbach2024non} results in excessive fragmentation.
Because searchers frequently utilize hundreds of addresses to circumvent Ethereum's nonce limitations, address-level mapping fails to capture the coherent behavior of a single underlying strategy.
While leveraging external labels offers an alternative~\cite{oz2024wins}, this method cannot guarantee exhaustive coverage of all trading activities.
Consequently, the field lacks a unified order flow definition that simultaneously enables an exhaustive evaluation of the landscape and the granularity essential for meaningful strategic analysis.

\noindent\textbf{RQ 1 (Sec.~\ref{subsec:bribe}):} \textit{How can we define and identify order flows in a manner that captures the coherent behavior of complex strategies while maintaining exhaustive coverage of the landscape?}

\smallskip\noindent\textbf{Gap 2: Incomplete Analysis of Exclusivity.} 
Recent studies have underscored the relationship between builder dominance and access to exclusive, or private, order flows (EOFs)~\cite{gupta2023centralizing,wang2024private,yang2025decentralization}.
Öz et al.~\cite{oz2024wins} characterize this dynamic as a ``chicken-and-egg'' dilemma: builders require EOFs to win blocks, while flow providers favor established builders with high inclusion rates.
However, current research lacks a systematic method for quantifying the degree of exclusivity across all order flows.
Developing such a metric is essential for empirically evaluating the extent of this dilemma and EOFs' overall impact.

\noindent\textbf{RQ 2.1 (Sec.~\ref{subsec:model} and~\ref{subsec:eofs}):} \textit{How can we quantitatively measure the exclusivity of all order flows?}

These results allow us to evaluate the overall importance of EOFs and test the ``chicken-and-egg'' hypothesis empirically:

\noindent\textbf{RQ 2.2 (Sec.~\ref{subsec:eofsperbuilder}):} \textit{To what extent do EOFs contribute to individual builder bids?}

\noindent\textbf{RQ 2.3 (Sec.~\ref{subsec:eofsperbuilder}):} \textit{Do non-dominant builders maintain their own EOFs, and if so, do EOFs constitute their primary source of revenue?}


\smallskip \noindent \textbf{Gap 3: Underexplored Non-atomic MEV Contracts.}
While Heimbach et al.~\cite{heimbach2024non} identified the existence of non-atomic arbitrage, their analysis was limited to a small set of instances. 
Due to the absence of a scalable classification method, the broader landscape of non-atomic MEV remains opaque. 

\noindent\textbf{RQ 3.1 (Sec.~\ref{subsec:random_forest_model} and~\ref{subsec:further_detection}):} \textit{Can we develop a method to identify all non-atomic MEV order flows?}

\noindent\textbf{RQ 3.2 (Sec.~\ref{subsec:non-atomic_mev_strategy_by_pool}):} \textit{Which non-atomic MEV flows are exclusive, and what magnitude of revenue do they generate for builders?}

\noindent\textbf{RQ 3.3 (Sec.~\ref{subsec:non-atomic_mev_strategy_by_pool}):} \textit{How is the revenue captured by builders distributed across non-atomic MEV flows?}

\noindent\textbf{RQ 3.4 (Sec.~\ref{subsec:non-atomic_mev_strategy_by_pool})}: \textit{Can we identify a correlation between this revenue and the number of the involved pools?} 

\smallskip \noindent \textbf{Gap 4: Lack of Temporal Analysis.}
Current research often measures the builder market as a static snapshot, yet the ecosystem has undergone substantial structural changes over the past two years.
Existing studies fail to capture the evolutionary dynamics of value extraction mechanisms and EOFs. This lack of a longitudinal perspective makes it difficult to resolve the causal factors behind market concentration.


\noindent\textbf{RQ 4.1 (Sec.~\ref{subsec:4phases}):} \textit{Can we quantitatively map the progression of builder market centralization?}

\noindent\textbf{RQ 4.2 (Sec.~\ref{subsec:shiftingValueExtraction}):} \textit{How have value extraction mechanisms and their relative profitability shifted throughout the study period?}

The following two questions evaluate the specific contributions of EOFs to the observed centralization:

\noindent\textbf{RQ 4.3 (Sec.~\ref{subsec:causality}):} \textit{Are EOFs a consistent prerequisite for builder success across different market regimes?}

\noindent\textbf{RQ 4.4 (Sec.~\ref{subsec:causality}):} \textit{Does an increase in bribe from EOFs correlate directly with an increase in market share?}

\smallskip\noindent\textbf{Gap 5: Root Causes of Centralization.} While various factors contributing to builder concentration have been theorized, current research lacks a meta-analysis that synthesizes longitudinal empirical data to isolate the primary drivers of this trend.
We hope our empirical results offer preliminary insights into the question:

\noindent\textbf{RQ 5 (Sec.~\ref{sec:discussion}):} \textit{Are EOFs and MEV the primary root causes of builder market centralization?}
\section{Data Collection and Preprocessing}
\label{sec:datacollection}
We begin our technical investigation by collecting historical Ethereum blockchain data and extracting swap-related transactions (Sec.~\ref{subsec:datacollection}).
We then formally define order flow and builder revenue (Sec.~\ref{subsec:bribe}).
Finally, we present a manual annotation protocol for the most lucrative contracts to establish the ground truth for classifying value extraction mechanisms (Sec.~\ref{subsec:label_contract}).



\subsection{Collecting and Processing Transaction Data}
\label{subsec:datacollection}
\textbf{Extracting Swap Transactions.}
Our study utilizes Ethereum transaction data spanning from September 1, 2023, to August 31, 2025, in the UTC timezone.
To facilitate efficient querying, we first deploy an Erigon archive node~\cite{erigon} to synchronize all on-chain data and historical states.
We then extract raw block and transaction data from block \num{18037988} through \num{23264565}.
This dataset encompasses \num{5226578} blocks and \num{889227817} transactions, which generated \num{503853.69} ETH in priority tips and direct bribes.


Transactions are categorized as either simple token transfers between accounts, referred to as \textit{non-swap} transactions, or \textit{swap} transactions, which involve the exchange of distinct token types.
Our analysis focuses on swap transactions because they are the builders' primary source of income.
Within our dataset, we identified \num{152466013} swap transactions.
Although these represent only 17.15\% of the total transaction count, they account for 78.97\% of the builders' revenue, totaling \num{397903.86} ETH.


\smallskip\noindent\textbf{Labeling Swap Transactions.}
We analyze these swap transactions across leading DEXes~\cite{defillama}, including UniswapV2, UniswapV3, Sushiswap, Curve, Balancer, and other platforms with compatible data formats such as PancakeSwap.
By implementing a custom swap extractor, we retrieve granular data including trading pair contract addresses, swap amounts, input/output token addresses, and sender/recipient identities.
To distinguish between public and private order flows, we compare our on-chain dataset with mempool data from the Mempool Guru project~\cite{yang2022sok}.
Any transaction observed on-chain but absent from the mempool dataset is labeled as private.
This categorization is vital because highly profitable MEV transactions are often submitted privately to builders to prevent competitors from front-running trades or replicating strategies.


\subsection{Processing Order Flow and Builder Data}
\label{subsec:bribe}
\textbf{Defining an Order Flow.}
To address \textbf{RQ 1}, we require a definition that overcomes the granularity-coverage dilemma by identifying a stable unit of strategic activity.
We observe that the destination contract directly invoked by a transaction serves as a robust indicator of its underlying purpose.
While searchers rotate sender addresses to circumvent nonce limitations, each underlying strategy remains bound to a specific destination contract; similarly, transactions invoking public protocols typically represent standard user trades.
We therefore define an order flow as the set of all transactions invoking the same destination contract.
An order flow's contribution to the gross income of one or more builders is defined as its \emph{bribe}, which we calculate as the aggregate bribe of all constituent transactions. 
This approach bypasses the fragmentation of address-level tracking while providing far greater granularity than simple visibility-based classifications.
Furthermore, it ensures exhaustive coverage by allowing us to categorize any swap transaction into a specific flow without prior knowledge of the participating searcher or builder.
By applying this methodology, we identify \num{164249} distinct order flows within our dataset.
Henceforth, we do not distinguish between an order flow and its corresponding contract when the context permits.

\smallskip\noindent\textbf{Identifying Builders.}
Leveraging publicly available sources~\cite{titanbuilder,ratedbuilders} and prior studies~\cite{heimbach2024non,heimbach2023ethereum}, we identify 59 distinct builders associated with 102 addresses.
There are more addresses than builders because a builder may utilize multiple receiving addresses.
For example, \textit{bobTheBuilder.xyz} maintains the largest footprint in our study with nine addresses.

We classify builders into three categories based on their peak historical weekly market dominance. Builders that have achieved a market share exceeding 50\% are defined as \emph{dominant builders}, including Titan and Beaverbuild.
Those that have never surpassed the 50\% threshold but have reached a share greater than 10\% are categorized as \emph{influential builders}, such as Lido, rsync-builder, build\-er0x69, Flashbots, jetbldr.xyz, and penguinbuild.org.
Finally, we designate all remaining entities as \emph{niche builders} if their historical market share has never reached the 10\% mark.

\smallskip\noindent\textbf{Quantifying Builders' Revenue.}
Calculating a builder's gross income solely by the difference in their balance before and after block execution can introduce inaccuracies because builders may initiate outgoing transfers within the block.
To mitigate this, we measure the builder's balance change $d_t$ for each transaction $t$, accounting for both priority tips and direct bribes as described in Sec.~\ref{subsec:pbs}.
We then aggregate non-negative balance changes to determine the total revenue $R_B$ in a block $B$ as $R_B = \sum_{{t}\in B} \max(d_t, 0)$.
Finally, the builder’s net profit for block $B$ is calculated by subtracting the bid $b_B$ paid to the proposer: $R_B - b_B$.

\subsection{Labeling Trading Contracts}
\label{subsec:label_contract}





To establish a ground truth for identifying non-atomic MEV flows in Sec.~\ref{sec:non-atomic}, we manually label the top 210 trading contracts by their bribes.
These contracts collectively account for 95\% of all builders' revenue between September 1, 2023, and August 31, 2025.

\smallskip\noindent\textbf{Collecting Public Labels.}
We begin by describing the dataset compiled for this labeling task.
Most public DeFi \emph{protocol contracts} are open-source and indexed on Etherscan~\cite{etherscan} by developers or community contributors.
From Etherscan’s labeled dataset, we collected DeFi protocol-related addresses.
To identify \emph{atomic MEV contracts}, we utilize labels from the ZeroMEV API~\cite{zeromev}, which categorizes transactions into sandwich attacks, atomic arbitrage, and liquidations.
Note that due to an unresolved issue, MEV labels are unavailable for the period between March 15 and March 16, 2024.
Finally, we incorporate findings from Heimbach et al.~\cite{heimbach2024non}, who identified thirteen \emph{non-atomic MEV contracts} in early 2024.

\smallskip\noindent\textbf{Classifying Top Trading Contracts.}
We implement a multi-stage labeling process to categorize high-bribe contracts based on their operational characteristics:

\begin{enumerate}
\item \textbf{Automated API Classification.} We prioritize labels from established sources. A contract is classified as atomic MEV if the ZeroMEV API labels more than half of its transactions as such. Similarly, any contract carrying a protocol label on Etherscan is categorized as a protocol.
\item \textbf{Deployer Linkage.} We extend protocol classification to include contracts deployed by known protocol developers. This heuristic assumes that deployer addresses for major protocols are managed with high security and are not repurposed for deploying unrelated searcher strategies.
\item \textbf{Forensic Sampling.} For unclassified contracts, we randomly sample ten transactions across different periods for manual inspection.
Inspecting with EigenTx~\cite{eigenphi} and Tenderly~\cite{tenderly}, we classify a contract as atomic MEV if the sampled transactions consistently generate riskless profit within a single block, either in isolation or through a bundle.
\item \textbf{Non-Atomic Identification.} Contracts are classified as non-atomic MEV if their transactions consistently exhibit unidirectional trading---buying or selling---without a corresponding reverse trade in the same block. Unlike the approach in~\cite{heimbach2024non}, we do not restrict the token set, as non-atomic strategies frequently target high-volatility assets recently listed on centralized exchanges.
\item \textbf{Residual Classification.} Contracts that do not align with the above criteria are labeled as \textit{other}. This category includes wash trading contracts, such as those identified in~\cite{washtrader1, washtrader2}, which are characterized by high transaction density in narrow timeframes involving a single token to inflate volume.
\end{enumerate}

By applying this methodology, we identified 113 atomic MEV contracts, 64 non-atomic MEV contracts, 30 protocol contracts, and 3 other contracts.
This annotated dataset provides the essential ground truth for our later analysis.


\section{Identifying and Analyzing\\Exclusive Order Flows}
\label{sec:eof}

This section evaluates the prevalence and impact of EOFs across the builder landscape.
We first define an exclusivity metric in Sec.~\ref{subsec:model} to identify EOFs while accounting for statistical biases.
In Sec.~\ref{subsec:eofs}, we apply this metric to our transaction dataset to characterize all EOFs.
Finally, Sec.~\ref{subsec:eofsperbuilder} analyzes the revenue contribution of EOFs to individual builders, demonstrating that access to EOFs is not limited to dominant builders: smaller builders often leverage niche EOFs to maintain operational viability.


\subsection{Quantifying Exclusivity}
\label{subsec:model}
While existing studies provide foundational insights into order flow visibility, many of them~\cite{heimbach2024non, oz2024wins, yang2025decentralization} rely heavily on simple correlation analyses that fail to account for inherent statistical biases.
Developing a universally applicable exclusivity metric therefore requires a framework that explicitly mitigates two primary confounding factors.
First, an order flow confirmed primarily by a dominant builder is not necessarily exclusive; a builder with a significant market share is statistically more likely to capture a higher proportion of public order flows by sheer volume---a phenomenon called \textit{endogeneity bias}.
Second, we must address the \textit{sample size confound}, which refers to statistical noise introduced by small order flows.
A low-frequency flow or one active only during a short period of market dominance may appear exclusive to a leading builder due to stochastic error rather than a deliberate strategic preference.



To address these challenges, we define an order flow's exclusivity within a discrete time window---e.g., one week---as the product of two components.
The first component measures the Kullback-Leibler (KL) divergence between the order flow's bribe distribution across builders and the builders' overall market share distribution.
This divergence directly accounts for endogeneity bias by penalizing flows that merely track market-wide dominance.
The second component weights this divergence by the square root of the flow's total bribe volume, thereby mitigating the sample size confound associated with low transaction frequencies.
We use the square root to prevent high-volume flows from disproportionately skewing the metric.
Finally, we aggregate these weekly scores to determine an order flow's overall exclusivity.
This longitudinal aggregation accounts for dynamic shifts in exclusivity---such as when a flow is sent exclusively to one builder for a specific duration before migrating to another---while further reducing noise from short-lived flows.


\begin{definition}[Order Flow Exclusivity]
Consider a market with $N$ builders and a set of temporal epochs $\mathcal{T}$.
For any epoch $t \in \mathcal{T}$, let $Q_{t,i}$ denote the number of blocks built by builder $i$, and $Q_t = \sum_{i=1}^N Q_{t,i}$ represent the total block number.
The aggregate market structure is defined by the distribution $\mathcal{S}_t = \{s_{t,i}\}_{i=1}^N$, where $s_{t,i} = Q_{t,i} / Q_t$ is the market share of builder $i$.

For a specific order flow $j$, let $R_{t, j \rightarrow i}$ be the revenue (bribe) received by builder $i$ from flow $j$ during epoch $t$.
The total revenue for flow $j$ is $R_{t,j} = \sum_{i=1}^N R_{t, j \rightarrow i}$. 
The routing preference of flow $j$ is represented by the distribution $\mathcal{P}_{t,j} = \{p_{t, j \rightarrow i}\}_{i=1}^N$, where $p_{t, j \rightarrow i} = R_{t, j \rightarrow i} / R_{t,j}$.

To quantify the deviation of an order flow's routing preference from the baseline market structure, we employ the Kullback-Leibler divergence:
\begin{equation}
    D_{\mathrm{KL}}(\mathcal{P}_{t,j} \| \mathcal{S}_t) = \sum_{i\in \{1,\dots,N\}} p_{t,j\rightarrow i} \log\left(\frac{p_{t,j\rightarrow i}}{s_{t,i}}\right) \,.
\end{equation}
The \emph{exclusivity} metric $\mathcal{E}(j)$ for order flow $j$ is defined as the cumulative volume-weighted divergence over all active epochs:
\begin{equation}
\mathcal{E}(j) = \sum_{t \in \mathcal{T}} \left( D_{\mathrm{KL}}(\mathcal{P}_{t,j} \| \mathcal{S}_t) \times \sqrt{R_{t,j}} \right) \,.
\end{equation}
\end{definition}

This formulation ensures that order flows exhibiting high routing persistence toward specific builders---disproportionate to those builders' market shares---receive a high exclusivity score.
By aggregating all epochs and incorporating the volume-weighting term $\sqrt{R_{t,j}}$, the metric effectively filters out ephemeral and low-value flows that might otherwise appear exclusive due to statistical noise.

\subsection{Empirical Identification of EOFs}
\label{subsec:eofs}

\begin{figure}[t]
    \includegraphics[width=\columnwidth, trim=0 12 0 0]{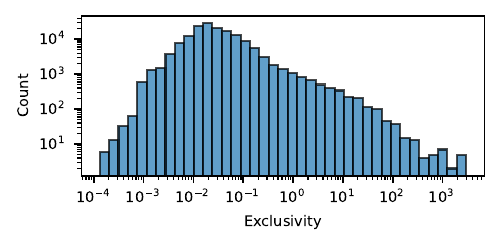}
    \caption{Distribution of order flow exclusivity $\mathcal{E}$.}
    \label{fig:eof_dist}
\end{figure}

\begin{figure}[t]
    \includegraphics[width=\columnwidth, trim=0 5 0 0]{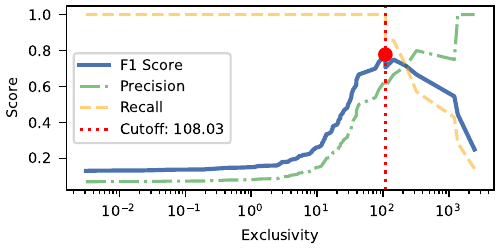}
    \caption{Threshold optimization via F1-score maximization.}
    \label{fig:exclusivity_threshold}
\end{figure}

\begin{figure*}[t]
    \centering
    \includegraphics[width=\linewidth, trim=0 12 0 0]{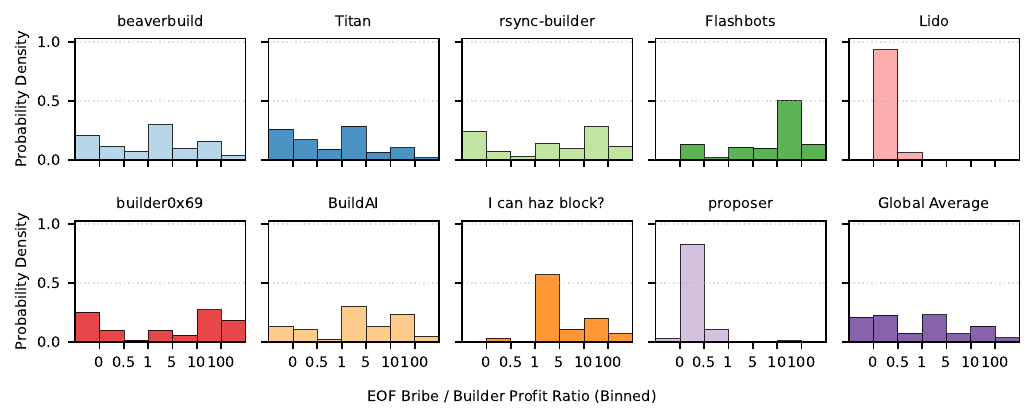}
    \caption{Distribution of EOF dependency ratios across eight builders, the proposer category, and a global average. The $x$-axis uses \textbf{ordinal bins} with fixed breakpoints ($0, 0.5, 1, 10, 100$) to visualize distinct operational modes.}
    \label{fig:eof_ratio_dist}
\end{figure*}

To address \textbf{RQ 2.1}, we compute the exclusivity for all identified contracts and analyze their distribution.
As illustrated by the log-log histogram in Fig.~\ref{fig:eof_dist}, their exclusivity exhibits a pronounced heavy-tailed distribution.
This statistical pattern validates our metric design: while the vast majority of flows cluster at the lower end of the spectrum---representing stochastic noise or public protocols---a distinct set of outliers in the rightmost tail clearly identifies EOFs.
Such a separation reinforces the hypothesis that exclusive routing strategies are statistically distinguishable from standard transaction traffic, allowing for the precise isolation of EOFs.


To establish a binary classification for exclusivity, we construct a ground-truth dataset comprising 175 contracts, including known EOFs cited in literature~\cite{daian2020flash, heimbach2024non, oz2024wins, yang2025decentralization} and public protocols labeled on Etherscan.
We evaluate the classification performance across varying thresholds and identify an optimal cutoff at $\tau = 108.03$, yielding a maximum F1-score of 0.78, as illustrated in Fig.~\ref{fig:exclusivity_threshold}.
Consequently, we classify any contract $j$ with an exclusivity score $\mathcal{E}(j) > 108.03$ as an exclusive order flow.


Applying this threshold yielded an initial set of 76 candidate EOFs.
Through manual verification, we identified one false positive: the Uniswap Router---the primary gateway for users and external applications to interact with Uniswap.
Due to its substantial accumulated bribe volume (\num{11810.02} ETH), the router achieved a score of 108.07, marginally exceeding our threshold of 108.03.
Given that Uniswap functions as a canonical public protocol where transactions are predominantly public, we excluded it from the final EOF dataset to ensure the rigor of our subsequent analysis.

Consequently, we identified a total of 75 EOFs.
It is noteworthy that while nine EOFs have been documented in prior literature, only seven were active during our observation period.
Therefore, the remaining 68 identified EOFs are previously unreported.
An overview of the top 20 flows, ranked by exclusivity, is provided in Table~\ref{tab:eof_list} (Appendix~\ref{app:eof_list}).
From an economic perspective, EOFs account for \num{280554.89} ETH in bribes, representing 70.53\% of total trading-related revenue during our observation period.
Notably, newly identified EOFs contribute approximately \num{98077.78} ETH, or 34.96\% of the EOF bribe.
This significant proportion of ``hidden'' exclusive flows cannot be attributed merely to our expanded observation window.
In fact, a direct comparison within a common observation window (September 1, 2023, to March 31, 2024) reveals that 37 EOFs remained active yet undetected by existing studies~\cite{daian2020flash, heimbach2024non, oz2024wins, yang2025decentralization}, highlighting the superior sensitivity of our metric.

\subsection{Builder Dependencies on EOFs}
\label{subsec:eofsperbuilder}

\textbf{Quantifying Dependency.} To address the relationship between market dominance and order flow exclusivity raised in \textbf{RQ 2.2}, we introduce the \emph{EOF Dependency Ratio (EDR)}.
For a given block, the EDR is defined as the ratio of EOF-derived bribes to the builder’s net profit.
This metric serves as a robust indicator of a builder's profit strategy and operational sustainability.
We partition the EDR into several intervals to characterize builder behavior.
Values in the $(-\infty,0)$ range signify negative-profit blocks, indicating that the builder is either capturing off-chain revenue or intentionally subsidizing searchers to defend market share.
The $[0,0.5)$ range suggests that EOFs constitute a minority of the builder's profit, a profile typical for builders primarily utilizing public order flows.
Conversely, a high EDR reveals a structural reliance on exclusive flow.
In the $[0.5,1)$ range, the builder exhibits a majority reliance on EOFs, while values in $[1,+\infty)$ denote cases where the builder's net profit is only a fraction of its total EOF revenue.
In this final scenario, the builder's competitive revenue is insufficient to sustain a winning bid to the proposer without EOF access, forcing the builder to relay all non-EOF revenue and a portion of its exclusive bribes as a tip to the proposer.


\smallskip\noindent\textbf{Empirical Distribution of EDR.}
Figure~\ref{fig:eof_ratio_dist} illustrates the EDR distribution across eight builders, the proposer category, and a global average for reference.
Blocks that cannot be attributed to a known builder are labeled as ``proposer'' blocks, as these are typically constructed directly by the proposer, bypassing the MEV-Boost architecture.
The x-axis is partitioned into six discrete intervals based on their economic implications. 

Consistent with EDR's design as an indicator of builder behavior, the majority of ``proposer'' blocks fall within the $[0,0.5)$ range. 
A small fraction (approximately 1\%) exhibit a negative EDR, likely representing smaller builders not yet identified in our dataset.

The \emph{dominant builders}, Titan and Beaverbuild, show both high EOF reliance and a significant proportion of negative-EDR blocks. 
This latter suggests a strategy of searcher subsidization to maintain high block-winning frequencies, or potentially the receipt of off-chain reimbursements.

Among \emph{influential builders}, rsync-builder and builder0x69 similarly demonstrate high EOF reliance and frequent negative EDRs. 
Flashbots displays a distinct structural dependency on EOF bribes, with half of its blocks concentrated in the $[10,100)$ category. 
In contrast, Lido operates predominantly with an EDR of 0, indicating minimal EOF integration and negligible subsidization.

Finally, although most \emph{niche builders} possess no EOFs, exclusivity often manifests through highly selective relationships between specific order flow sources and these builders, directly addressing \textbf{RQ 2.3}.
For instance, an atomic arbitrage contract (\texttt{0x6980...bdd0}) generated \num{5715.40} ETH in bribes, with 97.4\% of volume routed exclusively to ``I can haz block?''.
Similarly, an atomic arbitrage contract (\texttt{0x6454...4bfa}) contributed \num{4361.08} ETH with 94.7\% of transactions directed to a single builder address (\texttt{0x3bee...436}), yielding a high exclusivity score of $\mathcal{E} \approx 2697.17$.
These cases demonstrate that small- to mid-sized builders can maintain operational viability by serving as specialized, trusted execution channels for high-risk strategies.

\section{Identifying and Analyzing\\Non-Atomic MEV Order Flows}
\label{sec:non-atomic}
Despite their well-known contributions to builder revenue, non-atomic MEV contracts remain largely unexplored because their on-chain footprints often appear as standard token exchanges.
The pioneer work of Heimbach et al.~\cite{heimbach2024non} established five heuristics for detecting non-atomic arbitrage, achieving an accuracy range of 58\% to 90\% for known arbitrageurs.
However, their methodology identified only thirteen contracts, a result we believe significantly underestimates the true landscape due to three primary limitations.

First, their approach extracts features exclusively at the transaction level, thereby neglecting the rich statistical data and critical contextual information available at the contract level.
Second, by encoding all features as binary indicators, their model discards the nuanced information embedded in continuous values.
Third, their focus is restricted to a narrow set of assets, which overlooks the long tail of niche tokens and small-scale exchanges that often constitute the primary revenue source for some players.
Because of these rigidities, their method suffers from high false-negative rates and fails to capture the full spectrum of arbitrage activity.
Furthermore, since the transaction frequency of most order flows is lower than that of the top-tier contracts, it is infeasible to extend our manual annotation method (Sec.~\ref{subsec:label_contract}) to the entire market.

This motivates the supervised learning framework we propose in this section.
In Sec.~\ref{subsec:random_forest_model}, we construct a nine-dimensional feature vector for each contract, justify the selection of our machine learning algorithm, and detail the subsequent training process.
We then provide a comprehensive statistical overview of the identified non-atomic MEV flows in Sec.~\ref{subsec:further_detection} and evaluate their underlying strategies in Sec.~\ref{subsec:non-atomic_mev_strategy_by_pool}.

\subsection{Supervised Learning Framework}\label{subsec:random_forest_model}
\textbf{Feature Extraction.}
We represent each contract using a nine-dimensional feature vector categorized into two distinct groups.
The first group captures the average transaction-level behavior by aggregating metrics across its associated transactions.
These metrics include (1) the number of swap events triggered, (2) gas usage, representing computational cost, (3) priority tips, (4) the transaction's index within a block, (5) the frequency of MEV labels, and (6) the frequency of private labels.
While this group encompasses the heuristics proposed by Heimbach et al.~\cite{heimbach2024non}, we intentionally omit their ``established tokens'' constraint to reflect the reality that arbitrage occurs across the entire asset spectrum.
The second group describes contract-level characteristics: (1) the number of unique interacting addresses, (2) the average transaction count per sender, and (3) the total number of transactions executed by the contract.
Notably, all features are treated as continuous variables rather than binary indicators to preserve granular information.

To evaluate the efficacy of these features, we analyze our order flow database and present the resulting distributions in Fig.~\ref{fig_features} of Appendix~\ref{sec:appendix_features}.
We highlight the subset of 210 contracts manually annotated in Sec.~\ref{subsec:label_contract} to serve as a benchmark.
The visualization reveals clear distinctions between contract types, confirming that our selected feature set serves as a robust indicator for identifying value-extraction mechanisms.

\smallskip\noindent\textbf{Classification via Random Forest.}
We require a machine learning algorithm that offers strong performance despite the scarcity of ground-truth labels while allowing for calibrated model complexity to prevent overfitting and maintain interpretability.
To satisfy these requirements, we employ a Random Forest classifier.
As an ensemble method based on decision trees, Random Forest is inherently resilient to the severe overfitting common when training on small datasets.
The robustness of the model is further enhanced through majority voting across the forest, which mitigates the variance associated with individual trees.
Furthermore, because each tree partitions the feature space through discrete threshold-based rules, the model maintains a high degree of transparency and allows for clear visualization of the underlying classification logic.


\smallskip\noindent\textbf{Model Training and Evaluation.} We partition our set of 210 labeled contracts by reserving 30\% for an independent test set.
The remaining contracts are used to train an ensemble of seven decision trees using a cross-validation approach: the data is split into seven uniform subsets, with each tree using one subset for validation and the other six for training.
This procedure yields individual decision trees with an average accuracy of 91.61\% and a standard deviation of 0.0250.
The aggregated Random Forest achieves a classification accuracy of 92.06\% on the held-out test set.
The ensemble model demonstrates high classification certainty, with 84.1\% (53/63) of the test contracts achieving unanimous consensus (7 votes), while only a marginal fraction (3/63) fell into the borderline category with a narrow majority of 4 votes.

Feature importance analysis reveals that ``the average number of swap events per transaction'' is the most predictive metric for detecting non-atomic behavior, as these contracts typically exhibit exactly one swap event per transaction.
A representative decision tree is visualized in Fig.~\ref{fig:decision_tree} of Appendix~\ref{app:d_tree}.




\subsection{Empirical Discovery\\of Non-Atomic MEV Order Flows}
\label{subsec:further_detection}
\begin{figure*}[tb]
    \centering
    \includegraphics[width=\linewidth]{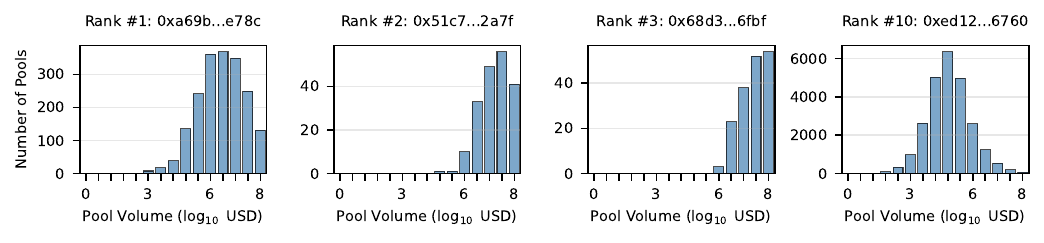}
    \caption{Distribution of the volume of trading pools exploited by top non-atomic order flows.}
    \label{fig:pool_distribution}
\end{figure*}

Building on the 210 manually annotated contracts, we leverage the supervised model to classify the remaining 790 order flows within the top \num{1000} bribe-contributing contracts.
We focus on this subset because these top \num{1000} contracts account for 98.63\% of the total trading-related builder revenue. 
Extending the classification to the full dataset of \num{164249} contracts would yield diminishing returns, as the long tail of order flows is characterized by sparse transaction history and negligible bribes, which degrade classification accuracy.

Our detection pipeline utilizes a two-stage classification process.
First, the Random Forest identifies 258 non-atomic MEV contracts.
Subsequently, we distinguish between atomic MEV and protocol-related contracts by extending the heuristics introduced in Sec.~\ref{subsec:label_contract}, specifically automated API classification, deployer linkage, and forensic sampling.
Of the remaining \num{532} contracts, this process identifies \num{113} as atomic MEV, \num{44} as protocol contracts, and \num{375} as miscellaneous order flows.

In total, our combined manual and automated efforts identify 322 non-atomic MEV contracts.
This set includes 316 newly discovered flows, as only six of the thirteen contracts previously identified by Heimbach et al. overlap with our observation window.
Economically, these non-atomic order flows generated \num{91483.07} ETH in bribes, representing 22.99\% of the trading-related builder revenue.
Critically, 39.84\% of this revenue originates from our newly discovered contracts, highlighting a significant expansion in the known scope of non-atomic MEV and the substantial role it plays in the builder market.

\subsection{Analyzing Non-Atomic MEV Order Flows}
\label{subsec:non-atomic_mev_strategy_by_pool}
\textbf{Exclusivity (RQ 3.2).}
We identify nine non-atomic flows that maintain exclusive collaborative relationships with dominant or influential builders.
Specifically, the 1st, 6th, 11th, 15th, and 17th largest flows by revenue contribution route exclusively to Beaverbuild;
the 3rd and 4th largest flows collaborate with Titan;
and the 2nd and 49th largest flows are affiliated with rsync-builder.
Collectively, these nine flows contribute \num{64986.14} ETH in bribes, accounting for 70.96\% of the total revenue generated by non-atomic MEV.
This extreme concentration suggests that the profitability of non-atomic strategies is tightly coupled with the specialized execution channels provided by the builder oligopoly.

\smallskip\noindent\textbf{Builder-Revenue Distribution (RQ 3.3).}
Our analysis identifies two distinct features characterizing the distribution of bribes across non-atomic order flows: a sharp structural concentration in the head and a super-heavy-tailed distribution in the tail. 
First, the revenue is heavily skewed toward a small number of dominant contracts; the top two non-atomic contracts alone capture 56\% of all non-atomic bribes, while the top ten account for 77\%.
Notably, only three of these top ten contracts (the 1st, 2nd, and 7th) were previously identified in~\cite{heimbach2024non}, whereas the remaining seven are first identified in this work.
This highlights the capacity of our supervised learning approach to uncover high-value flows that simpler heuristics overlook.
Detailed metrics for these leading contracts are presented in Table~\ref{tab:top_non_atomic} of Appendix~\ref{app:non-atomic}.

Second, the market exhibits a super-heavy-tailed distribution.
We modeled the relationship between contract rank and cumulative bribes using Maximum Likelihood Estimation to perform a power-law fit.
The resulting Kolmogorov-Smirnov statistic ($D = 0.098$) confirms a high goodness-of-fit.
The power-law exponent $\alpha = 1.47$ is particularly significant, as values in the range $1 < \alpha < 2$ characterize a distribution with theoretical infinite variance and an undefined mean in the limit.
Economically, this indicates that while a few ``black swan'' generators capture the majority of value, the market remains underpinned by a resilient and extensive tail of low-volume participants.

\smallskip\noindent\textbf{Correlation with Pool Complexity (RQ 3.4).}
We analyze the pool-level interaction patterns of each non-atomic flow to determine if their bribe contribution correlates with the breadth of pools exploited.
On average, each non-atomic contract interacts with \num{1247} pools, though the scale varies significantly, with a minimum of one pool observed for 51 contracts and a maximum of \num{45638} pools.
Our analysis reveals no linear correlation between the absolute number of pools utilized and total bribe.
For instance, within the top ten bribe-contributing contracts, the pool count spans more than two orders of magnitude, ranging from 209 to \num{30742}.

Instead, the bribe contribution appears more closely linked to \emph{pool volume}.
Each pool's volume, or size, is computed as the accumulated trading volume by all swap transactions in our measured period, measured in USD.
In general, higher-bribe contracts target pools with larger mean and median sizes.
Among the top ten flows, median pool sizes range from \num{131} thousand to \num{186.6} million USD, while mean sizes range from \num{14.8} million to \num{2.81} billion USD.
Figure~\ref{fig:pool_distribution} depicts the distribution of exploited trading volume for the top five flows.
However, notable outliers challenge the assumption that volume alone drives success.
The contract ranked 5th interacts with \num{2366} pools---far more than its higher-revenue peers---yet targets pools with mean and median sizes an order of magnitude smaller.
Similarly, the top two contracts employ divergent strategies: contract \texttt{0xa69b...e78c} interacts with significantly more pools than \texttt{0x51c7...2a7f} but maintains a transaction volume approximately ten times smaller.
These findings suggest that non-atomic arbitrageurs can capture substantial profits either through high-volume targeting in mainstream pools or by aggregating smaller yields across a fragmented landscape of overlooked pools.

\section{Temporal Evolution of Builder Centralization}
\label{sec:mechanics}
\begin{figure*}[tb]
\centering
\includegraphics[width=\linewidth, trim=0 0 0 0]{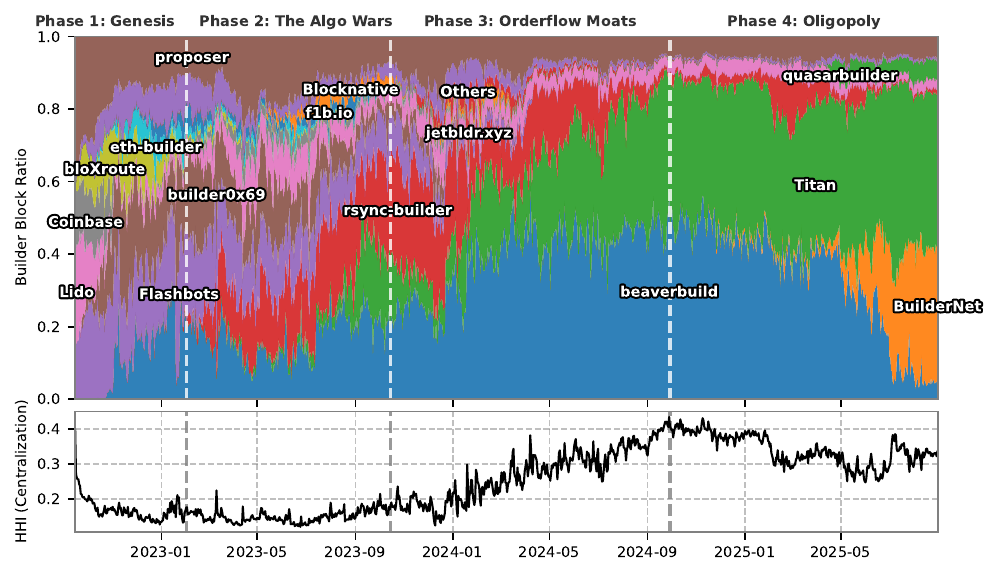}
\caption{Block ratio distribution of builders and Herfindahl-Hirschman index of the whole market.}
\label{fig_builders}
\end{figure*}

We now apply the dimensions of exclusivity and value extraction mechanisms to investigate the longitudinal development of the Ethereum builder market.

\subsection{Phases of Market Concentration\\and Their Revenue Composition}
\label{subsec:4phases}


\textbf{Quantifying Market Concentration via HHI.}
To assess the intensity of market centralization, we analyzed MEV-Boost block data from its inception on September 1, 2022, to August 31, 2025.
The longitudinal evolution of builder market shares is illustrated in Fig.~\ref{fig_builders}, which displays the block ratios of leading participants across this three-year period.
Beneath the market share distribution, we present the Herfindahl-Hirschman Index (HHI) as a continuous measure of concentration.
We compute the HHI by summing the squares of all builders’ fractional market shares on a weekly basis, where a higher index signifies a more centralized environment.
This metric provides a formal basis for categorizing the market's evolution into distinct phases, tracing the transition from an initially distributed landscape to the current state of high concentration dominated by a few major entities.

\smallskip\noindent\textbf{Phases of Market Concentration.}
Based on observed shifts in HHI and the entries and exits of prominent participants, we divide the investigated period into four phases:

\begin{description}
\item[Phase 1: Genesis] (September 2022 – January 2023) marks the introduction of the MEV-Boost framework by Flashbots.
During this initial stage, builders such as Lido and Coinbase leveraged their existing dominance in the proposer set to gain an early market advantage.
Concurrently, bloXroute capitalized on its physical relay infrastructure to minimize network latency, using these efficiencies to subsidize builder operations until its market share fell below 5\% in early 2023.

\item[Phase 2: Algorithm Wars] (January – October 2023) was characterized by the lowest fluctuations in market concentration and intense competition over MEV extraction efficiency.
Competitive advantage shifted from infrastructural scale to algorithmic sophistication, as builders sought to outperform proposers in identifying profitable bundles.
Notably, builder0x69 reached its peak market share during this phase before its eventual decline, while the emergence of Titan and rsync-builder signaled a transition toward more specialized extraction strategies.

\item[Phase 3: EOF Moats] (October 2023 – October 2024) experienced a sharp rise in the HHI, driven by the emergence of Beaverbuild as a dominant entity.
Amid this period, the relevance of low latency and algorithmic speed diminished as EOFs became the primary determinant of success. As we demonstrate in Sec.~\ref{subsec:causality}, the capture of EOFs significantly widened the revenue gap between top-tier builders and their competitors, establishing a formidable barrier to entry.

\item[Phase 4: Oligopoly] (October 2024 – August 2025) reflects the formation of a rigid market structure where Beaverbuild, Titan, and rsync-builder collectively controlled approximately 90\% of block production.
While new participants such as BuilderNet entered the market in 2025, our analysis reveals that BuilderNet's order flow distribution is nearly identical to that of Beaverbuild.
This suggests that the entrant primarily cannibalized the share of incumbents rather than restoring competition, indicating that the oligopolistic equilibrium remains persistent.
\end{description}

\begin{figure*}[tb]
\centering
\includegraphics[width=\linewidth, trim=0 0 0 0]{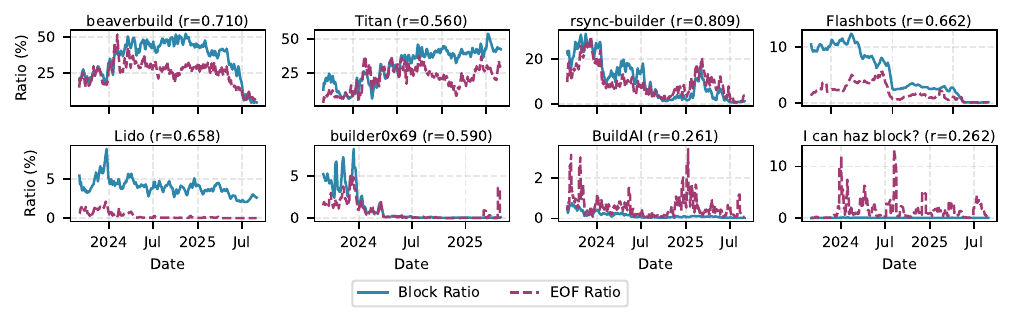}
\caption{Temporal dynamics of block market share (blue) and EOF ratio (red) across top builders, where the Pearson correlation coefficient ($r$) highlights the strong interdependence between EOF access and market dominance.}
\label{fig:builder_eof}
\end{figure*}

\subsection{Shifting Profiles of Value Extraction}\label{subsec:shiftingValueExtraction}
To address \textbf{RQ 4.2}, we analyze the changing composition of how three value extraction mechanisms contribute to builder revenue from Phase 2 to 4, noting that Phase 1 is excluded due to insufficient data.
Profits from protocol contracts declined steadily, dropping from an average of 42.4\% in Phase 2 to 20.7\% in Phase 4.
The average share of profits from atomic MEV contracts was 25.0\%, 32.6\%, and 22.5\% across phases, indicating a sharp decrease from Phase 3 to Phase 4.
In contrast, the share of profits generated by non-atomic MEV contracts rose from 13.2\% to 20.7\%, nearly matching the level of atomic MEV contracts by the end of the observation period.
The remaining share, attributable to miscellaneous flows, expanded from approximately 19.4\% in Phase 2 to 36.1\% in Phase 4.
A primary contributor to this growth is the proliferation of new atomic arbitrage strategies that the ZeroMEV API has not yet labeled, leading them to be classified as miscellaneous rather than atomic MEV.

The rising prominence of non-atomic flows signifies a fundamental increase in market entry barriers.
Unlike atomic MEV strategies, which are effectively risk-free due to transaction atomicity, non-atomic strategies such as CEX-DEX arbitrage require permanent capital allocation across disjoint venues.
These strategies introduce inventory risk and price-drift exposure during the settlement lag, creating a structural barrier that favors institutional-grade searchers equipped with deep liquidity and sophisticated off-chain risk management.
Evidence of this structural barrier is found in the highly concentrated bribe distribution among non-atomic flows analyzed in Sec.~\ref{subsec:non-atomic_mev_strategy_by_pool}.
Consequently, the increasing revenue contributions of non-atomic flows, combined with the exclusive builder relationships identified in Sec.~\ref{subsec:non-atomic_mev_strategy_by_pool}, serve to entrench the dominant market positions of the leading builders.



\subsection{Causal Dynamics of EOF-Driven Dominance}
\label{subsec:causality}

\textbf{Quantifying Linear Relationships via Pearson Correlation Coefficient.}
To investigate the causal dynamics of the ``chicken-and-egg'' dilemma, we analyze the structural dependencies between builder dominance and the bribe contributions from EOFs.
For each builder, we compute the Pearson correlation coefficient ($r$) between two continuous variables: the builder's daily block market share and the relative daily contribution of its EOF bribes to the total trading-related revenue across all builders.
This coefficient serves as a robust indicator of the causal relationship by measuring the strength and direction of linear dependencies.
Specifically, it determines whether increases in one variable correspond to proportional changes in the other.
While a correlation coefficient of $+1$ or $-1$ indicates a perfect linear relationship, a value of $0$ implies no linear correlation.

\smallskip\noindent\textbf{Empirical Analysis (RQ 4.3 and 4.4).}
Our longitudinal analysis reveals that the role of EOFs in sustaining builder dominance is highly regime-dependent. 
As illustrated in Fig.~\ref{fig:builder_eof}, Beaverbuild exhibits a stronger correlation with its EOFs than Titan, a relationship primarily driven by a structural shift in 2025. 
During this period, BuilderNet's absorption of Beaverbuild's order flows triggered a simultaneous decline in the incumbent's market share and EOF revenue. 
This synchronized downward trend mathematically reinforces the correlation.
Crucially, the relationship for both market leaders decouples during Phase 4.
This shift indicates that once an oligopoly is established, dominant builders no longer rely on marginal EOF income to preserve their market share; instead, established network effects naturally incentivize searchers to prioritize collaboration with these incumbents.

The dynamics differ significantly for \emph{influential builders} such as rsync-builder, Flashbots, and builder0x69.
These entities exhibit a consistently strong correlation between market share and EOF bribes; consequently, as their EOF bribes declined, their market shares followed a similar downward trajectory.
An exception is observed in Lido, which maintains a high correlation coefficient despite a low EOF ratio, suggesting the deployment of alternative strategies to stabilize its market position.

For \emph{niche builders}, the data suggests that EOFs alone are insufficient to catalyze growth.
Despite experiencing frequent spikes in EOF bribes, their market shares remain stagnant at low levels.

These patterns support the conclusion that while EOF acquisition was a prerequisite for success during Phase 3---as seen in the rise of Beaverbuild and Titan---it no longer guarantees market share expansion in the current oligopolistic regime.
This is most clearly demonstrated by the performance of ``I can haz block?'', where successful EOF capture failed to translate into broader market dominance.
\section{The Inevitable Centralization of PBS}
\label{sec:discussion}

\textbf{Revisiting the ``Chicken-and-Egg'' Dilemma.}
Our results corroborate the findings of Oz et al.~\cite{oz2024wins} regarding high entry barriers for new participants.
However, by analyzing all order flow data across two dimensions, we demonstrate that this barrier is not composed exclusively of EOFs.
While EOFs were instrumental in enabling Beaverbuild and Titan to establish their dominance during Phase 3 (Sec.~\ref{subsec:causality}), their current positions are also consolidated through searcher subsidies (Sec.~\ref{subsec:eofsperbuilder}) and broader network effects.
Once an oligopoly is formed, high-value EOFs are no longer sufficient for smaller builders to gain market share.
We therefore conclude that EOF is both a symptom of growing centralization and a catalyst for its acceleration, rather than the sole root cause.

\smallskip\noindent\textbf{EOF as the Primary Vehicle.} Despite this nuance, EOFs remain the primary vehicle of centralization, as evidenced by the significant rise in HHI during Phase 3 (Sec.~\ref{subsec:4phases}).
Historically, builders have competed via three advantages: proposer collaboration (Phase 1), algorithmic efficiency (Phase 2), and EOFs. 
Contrary to the analysis in Bahrani et al.~\cite{bahrani2024centralization}, collaboration with proposers has proved less lucrative than vertical integration with searchers, as the latter yields significantly higher returns. 
Furthermore, superior algorithms do not provide a durable moat because the transparent nature of the blockchain allows competitors to analyze and replicate any efficiency gains.

EOFs were selected by the market as the dominant vehicle for centralization because a barrier built on information asymmetry, economic alliances, and capital liquidity---specifically through non-atomic MEV (Sec.~\ref{subsec:shiftingValueExtraction})---is far more resilient than one based on reproducible code.
As noted by Wang et al.~\cite{wang2024private} and Wu et al.~\cite{wu2024competition}, EOFs allow builders to solidify market share with lower bids and reduced risk exposure.

\smallskip\noindent\textbf{The Actual Root Cause of Builder Centralization.} We argue that builder centralization is an emergent property of the PBS architecture itself, as it systematically violates three fundamental prerequisites of a competitive market.
First, the requirement for \emph{diminishing returns to scale} is absent. The market favors incumbents who leverage superior computational resources and comprehensive mempool views to reduce the marginal cost of block optimization toward zero.
Second, \emph{information symmetry} is fundamentally undermined. Dominant builders utilize deep searcher integrations and auxiliary off-chain data to maintain an informational advantage that new entrants cannot replicate.
Finally, the prerequisite of \emph{low entry barriers} is violated by immense ``cold start'' costs. While block-building software is open-source, the necessity of established access to exclusive searcher flows---particularly non-atomic flows---creates an insurmountable capital and social moat.
Furthermore, three design flaws within PBS accelerate these dynamics:
\begin{description}
\item[Ordinal Competition.] The system creates a ``winner-take-all'' trap where a bid only 0.001 ETH lower than the leader results in zero value, incentivizing extreme consolidation.
\item[Positive Feedback Loops.] Increased market share leads to higher EOF acquisition, which further increases market share, creating a self-reinforcing cycle of dominance.
\item[Loss of Consensus Defense.] By decoupling execution from consensus, PBS removed the natural anti-monopoly barrier provided by validator dispersion. This pushed block construction into a hyper-specialized financial arena where specialization inevitably breeds centralization.
\end{description}

\section{Related Work} \label{sec:related_work}
\noindent\textbf{MEV landscape and Mitigation.} Blockchain’s transparency enabled front-running attacks, letting attackers profit by intercepting transactions during initial coin offerings as early as 2017~\cite{eskandari2019sok}. 
The rapid rise of DeFi in 2021 amplified such opportunities, formalizing the concept of maximal extractable value~\cite{daian2020flash}. 
MEV strategies include sandwich attacks~\cite{9519421,heimbach2022eliminating,wang2022impact,9519469,varun2022mitigating}, cyclic arbitrage~\cite{wang2022cyclic,jin2022detecting}, liquidation~\cite{qin2021empirical,heimbach2021behavior}, price manipulation~\cite{10.1007/978-3-662-64322-8_1,wu2023defiranger,wang2024defiguard}, non-atomic arbitrage~\cite{heimbach2024non}, and general front-running~\cite{qin2023blockchain}. Pattern extraction techniques as well as deep learning methods have been applied to detect various MEV activities~\cite{li2023demystifying}.
Research on mitigating MEV focuses on three approaches: private channel auctions~\cite{heimbach2023ethereum}, order-fairness consensus to prevent front-running~\cite{kelkar2022order,10.1007/978-3-030-56877-1_16,10.1007/978-3-031-18283-9_15,10.1007/978-3-662-63958-0_17,10.1145/3419614.3423263,kelkar2023themis,heimbach2022sok,cachin2022quick,10177428,piet2023mevade}, and application-level designs incorporating privacy or batch execution~\cite{baum2021p2dex,baum2023eagle,ciampi2022fairmm}.

\smallskip\noindent\textbf{PBS Builder Centralization.} Prior work highlights centralization risks and profit-maximization tactics~\cite{gupta2023centralizing,wahrstatter2024blockchain}, such as builders’ bidding strategies~\cite{wu2024strategic,wu2024compete} and proposers delaying blocks~\cite{oz2023time,wu2024competition}. 
Flashbots, Beaverbuild, and Nethermind launched BuilderNet to curb builder centralization via shared order flows in a Trusted Execution Environment (TEE) hardware. 
By January 2026, BuilderNet produced 25.5\% of blocks~\cite{mevboost}. 
While aiming for privacy and neutrality, its TEE requirement raises entry barriers, and cooperative design removes internal competition, enabling builders to maximize profits without bidding against each other.

\section{Conclusion}
\label{sec:conclusion}

By rigorously measuring the dimensions of exclusivity and value extraction, this research demonstrates that EOFs and non-atomic MEV are not merely incidental features of the block-building landscape.
Instead, they function as both symptoms and catalysts of an entrenched oligopolistic trend.
Our mapping of the market's temporal evolution reveals a critical shift in the nature of dominance.
While EOFs served as the primary ``moat'' during the transition to a centralized state, current market leaders have effectively decoupled their market share from marginal EOF income.
Their positions are now solidified through network effects, searcher subsidies, and the high capital requirements of non-atomic MEV---institutional barriers that structurally disadvantage new entrants.
Ultimately, we conclude that builder centralization is an inevitable outcome of the current PBS architecture.
Addressing these dynamics will require a fundamental reassessment of how block construction interacts with consensus-layer security.

\appendix
\section*{Ethical Considerations}
This study does not involve human subjects or sensitive personal data. All experiments are conducted using publicly available blockchain data, and no ethical concerns are associated with the research methods employed.

\section*{Open Science}
Our code for this paper is available at \url{https://anonymous.4open.science/r/experiments-4677/}

\section*{Generative AI Usage}

We disclosed the use of Gemini (Google) to assist in several aspects of this study. The tool was utilized for editorial refinement (grammar and spelling checks) and to provide recommendations on statistical math tools suitable for our specific measurement requirements. Additionally, Gemini assisted in generating Python code for data visualization. All AI-generated outputs, including code and statistical advice, were manually verified and tested by the authors to ensure their accuracy and applicability to the experimental results.
\printbibliography
\appendix

\cleardoublepage
\appendix
\section*{Appendix}

\section{Complete List of High-Exclusivity Order Flow Sources}
\label{app:eof_list}

Table~\ref{tab:eof_list} presents 20 exclusive order flows with the highest exclusivity. 

\begin{table*}[b]
    \centering
    \caption{Summary of the top EOFs. Highlighted contracts correspond to EOFs previously identified in the literature.}
    \label{tab:eof_list}
    \begin{tabular}{llrrrr}
        \toprule
        \textbf{Contract Address} & \textbf{Label} & \textbf{Total Score} & \textbf{Total Bribe (ETH)} & \textbf{Avg. KL} & \textbf{Active Wks} \\
        \midrule
        \texttt{0x6980a47bee930a4584b09ee79ebe46484fbdbdd0} & atomic & 2,982.85 & 5,715.40 & 6.27 & 61 \\
        \texttt{0x64545160d28fd0e309277c02d6d73b3923cc4bfa} & atomic & 2,697.17 & 4,361.08 & 6.43 & 60 \\
        \textcolor{red}{\texttt{0x51c72848c68a965f66fa7a88855f9f7784502a7f}} & non-atomic & 2,428.80 & 24,543.74 & 1.58 & 105 \\
        \texttt{0xe08d97e151473a848c3d9ca3f323cb720472d015} & atomic & 2,427.48 & 3,904.45 & 7.20 & 75 \\
        \texttt{0x00000000003b3cc22af3ae1eac0440bcee416b40} & atomic & 2,027.75 & 4,283.83 & 3.18 & 105 \\
        \texttt{0xe20cd9377c204a27952f8b41075f0b8bd1ceec3d} & atomic & 1,408.58 & 1,432.99 & 7.20 & 35 \\
        \textcolor{red}{\texttt{0xa69babef1ca67a37ffaf7a485dfff3382056e78c}} & non-atomic & 1,354.43 & 27,161.48 & 0.86 & 105 \\
        \texttt{0xe8c060f8052e07423f71d445277c61ac5138a2e5} & atomic & 1,209.14 & 3,097.01 & 2.11 & 105 \\
        \textcolor{red}{\texttt{0x3328f7f4a1d1c57c35df56bbf0c9dcafca309c49}} & protocol & 1,199.35 & 56,607.91 & 0.54 & 92 \\
        \texttt{0x9d6b911199b891c55a93e4bc635bf59e33d002d8} & atomic & 1,198.15 & 4,378.46 & 6.41 & 15 \\
        \texttt{0x3aa228a80f50763045bdfc45012da124bd0a6809} & atomic & 1,029.79 & 1,181.19 & 7.20 & 30 \\
        \texttt{0xe2cd944360d75bf2c323e4ab8b64ea3bbc358181} & atomic & 961.27 & 880.70 & 3.97 & 83 \\
        \texttt{0x13cf8829d2eaf6a70a33c6b9bb67b6c2f5d135be} & atomic & 817.79 & 1,008.12 & 7.24 & 31 \\
        \texttt{0xa090fc409a9f25bf8e28257d42ef6904590c8984} & atomic & 811.62 & 1,484.41 & 6.61 & 21 \\
        \texttt{0xed0c66f41b1f588f61629ae7be979f004363f8a4} & non-atomic & 799.17 & 748.86 & 6.91 & 24 \\
        \texttt{0x80bf7db69556d9521c03461978b8fc731dbbd4e4} & atomic & 750.52 & 504.91 & 6.65 & 30 \\
        \texttt{0x97f0fa1e5bf44f0dbd8dc5ee5e0d017667ae0c34} & atomic & 682.33 & 643.78 & 8.32 & 15 \\
        \texttt{0x5e51328c0583094b76f28cfd532abc3d454fcfea} & atomic & 640.99 & 520.78 & 6.49 & 28 \\
        \texttt{0xb4b50bfa4c4c90dd25a4cbe695f8124d24ac6474} & atomic & 627.50 & 518.02 & 6.85 & 20 \\
        \texttt{0xee14d52f7544f84748eea641b9b616bd65aab073} & atomic & 425.14 & 388.65 & 3.84 & 41 \\
        \bottomrule
    \end{tabular}
\end{table*}

\section{Non-Atomic MEV Order Flow}\label{app:non-atomic}

Table \ref{tab:top_non_atomic} lists the top 20 non-atomic order flows ranked by total bribes. The data reveals that high-value non-atomic arbitrage is heavily dependent on EOF, and confirms our model's ability to identify significant actors missed by previous studies.

\begin{table*}[t]
    \centering
    \caption{Top bribe non-atomic: analysis of EOF status and known non-atomic labels.}
    \label{tab:top_non_atomic}
    \begin{tabular}{lrrr}
        \toprule
        \textbf{Contract Address} & \textbf{Bribe (ETH)} & \textbf{Is EOF} & \textbf{Is Known in \cite{heimbach2024non}} \\
        \midrule
        \texttt{0xa69babef1ca67a37ffaf7a485dfff3382056e78c} & 27,161.48 & True & True \\
        \texttt{0x51c72848c68a965f66fa7a88855f9f7784502a7f} & 24,543.74 & True & True \\
        \texttt{0x68d3a973e7272eb388022a5c6518d9b2a2e66fbf} & 6,036.68 & True & False \\
        \texttt{0x6f1cdbbb4d53d226cf4b917bf768b94acbab6168} & 2,693.44 & True & False \\
        \texttt{0xc6fecdf760af24095cded954de7d81ab49f8bae1} & 2,300.64 & False & False \\
        \texttt{0x5050e08626c499411b5d0e0b5af0e83d3fd82edf} & 1,866.31 & True & False \\
        \texttt{0x98c3d3183c4b8a650614ad179a1a98be0a8d6b8e} & 1,769.75 & False & True \\
        \texttt{0x5ddf30555ee9545c8982626b7e3b6f70e5c2635f} & 1,603.33 & False & False \\
        \texttt{0xfbeedcfe378866dab6abbafd8b2986f5c1768737} & 1,588.02 & False & False \\
        \texttt{0xed12310d5a37326e6506209c4838146950166760} & 1,283.63 & False & False \\
        \texttt{0x1bf621aa9cee3f6154881c25041bb39aed4ca7cc} & 1,051.81 & True & False \\
        \texttt{0xf3de3c0d654fda23dad170f0f320a92172509127} & 1,030.98 & False & False \\
        \texttt{0x738e79fbc9010521763944ddf13aad7f61502221} & 1,004.52 & False & False \\
        \texttt{0x6719c6ebf80d6499ca9ce170cda72beb3f1d1a54} & 782.14 & False & False \\
        \texttt{0xe8cfad4c75a5e1caf939fd80afcf837dde340a69} & 775.58 & True & True \\
        \texttt{0x99b1817acb40e76c309e26b2face9da9eff55317} & 774.18 & False & False \\
        \texttt{0x000000000dfde7deaf24138722987c9a6991e2d4} & 711.11 & True & True \\
        \texttt{0x73a8a6f5d9762ea5f1de193ec19cdf476c7e86b1} & 705.45 & False & False \\
        \texttt{0x12ff0e28318e53a6f91d42cf607963076af6c03f} & 665.20 & False & False \\
        \texttt{0x360e051a25ca6decd2f0e91ea4c179a96c0e565e} & 599.83 & False & False \\
        \bottomrule
    \end{tabular}
\end{table*}

\section{Non-Atomic Features}
\label{sec:appendix_features}
We evaluate a dataset of \num{164249} contracts using the extracted features and obtain the distribution of contracts according to the metrics depicted in Fig.~\ref{fig_features}. 

\begin{center}
\begin{minipage}{\textwidth}
    \centering
    \includegraphics[width=\linewidth, trim=0 0 0 0, clip]{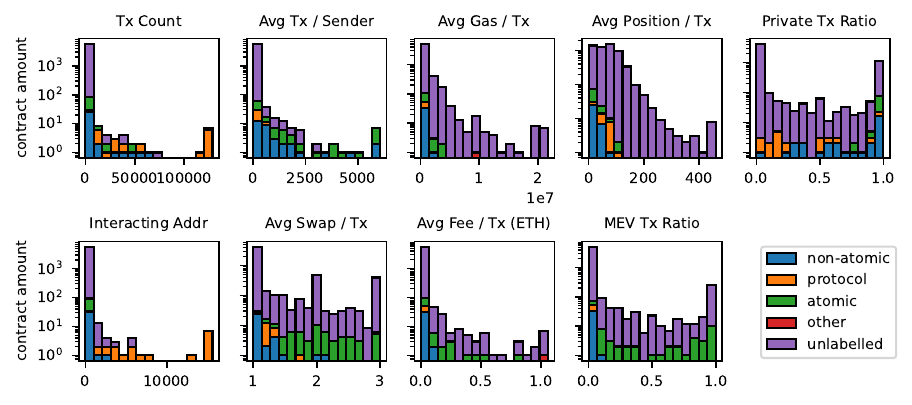}
    \captionof{figure}{Distribution of all order flows across different feature dimensions, including those with manual labels.}
    \label{fig_features}
\end{minipage}
\end{center}

\section{Known EOFs}
Table~\ref{tab:full_entity_mapping} presents a list of nine known EOFs from previous studies. 
\begin{table*}[h] 
    \centering
    \caption{List of nine known EOFs from previous studies}
    \label{tab:full_entity_mapping}
    \footnotesize
    \begin{tabular}{lllc}
        \toprule
        \textbf{Contract Address} & \textbf{Label} & \textbf{Description} & \textbf{Ref.} \\
        \midrule
        \texttt{0x3328f7f4a1d1c57c35df56bbf0c9dcafca309c49} & protocol & BananaGun & \cite{yang2025decentralization,oz2024wins} \\
        \texttt{0x51c72848c68a965f66fa7a88855f9f7784502a7f} & non-atomic & Rsyncsearcher & \cite{heimbach2024non,oz2024wins} \\
        \texttt{0xa69babef1ca67a37ffaf7a485dfff3382056e78c} & non-atomic & Beaversearcher & \cite{heimbach2024non,oz2024wins} \\
        \texttt{0x6b75d8af000000e20b7a7ddf000ba900b4009a80} & atomic & Jaredfromsubway & \cite{yang2025decentralization,oz2024wins} \\
        \texttt{0xdb5889e35e379ef0498aae126fc2cce1fbd23216} & protocol & BananaGun & \cite{yang2025decentralization,oz2024wins} \\
        \texttt{0x80a64c6d7f12c47b7c66c5b4e20e72bc1fcd5d9e} & protocol & Maestro & \cite{yang2025decentralization,oz2024wins} \\
        \texttt{0xf8b721bff6bf7095a0e10791ce8f998baa254fd0} & non-atomic & Mantasearcher & \cite{heimbach2024non} \\
        \texttt{0x57c1e0c2adf6eecdb135bcf9ec5f23b319be2c94} & non-atomic & builder1-searcher & \cite{heimbach2024non} \\
        \texttt{0xe8cfad4c75a5e1caf939fd80afcf837dde340a69} & non-atomic & searcher1 & \cite{heimbach2024non} \\
        \bottomrule
    \end{tabular}
\end{table*}

\section{A Representative Decision Tree}
\label{app:d_tree}
\begin{figure*}[t]
    \centering
    \includegraphics[width=\linewidth, trim = 0 0 30 0]{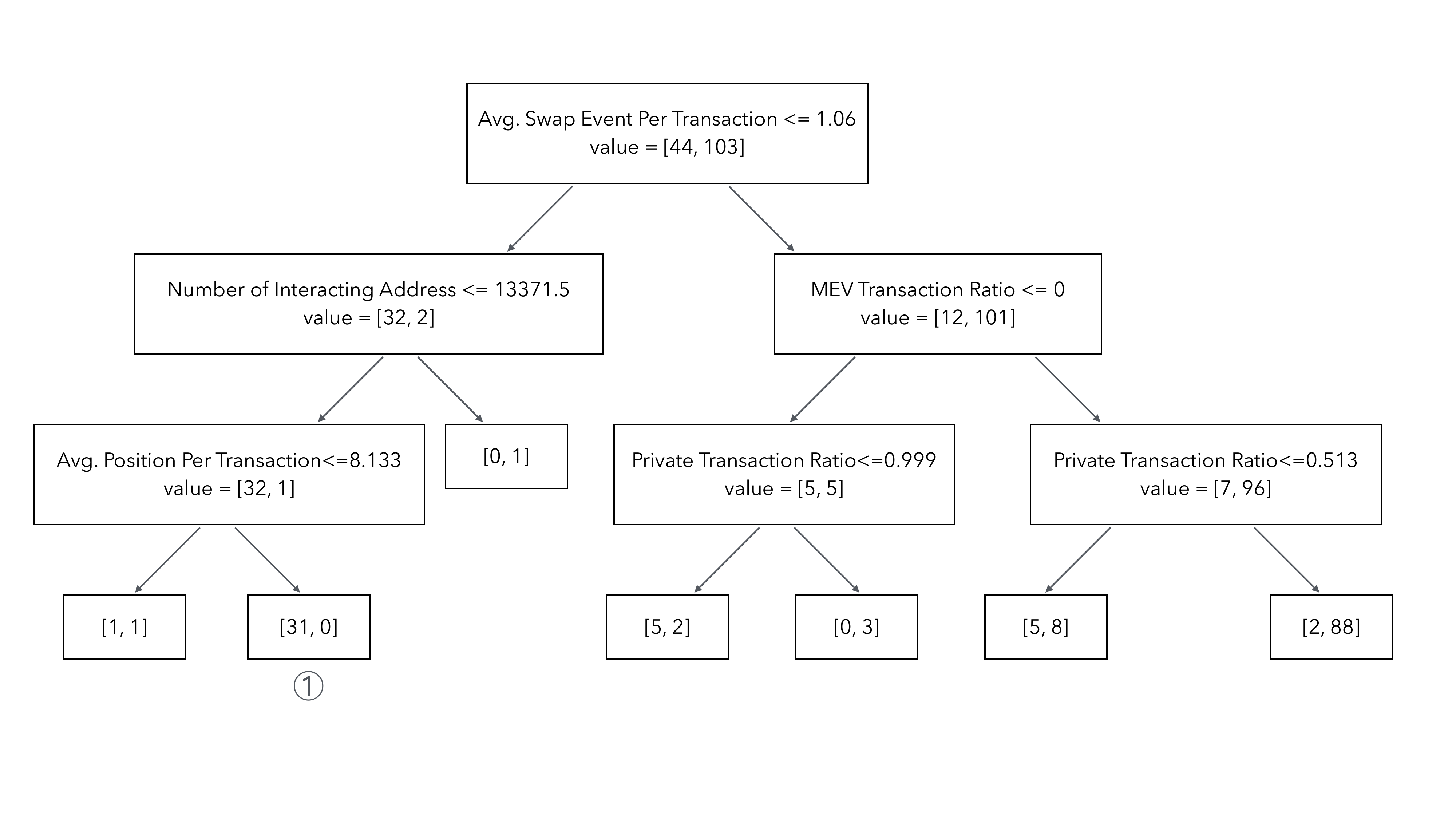}
    \caption{A Representative Decision Tree from the Random Forest. The first element of the value represents the number of non-atomic samples, and the second element represents the number of other samples. From \textcircled{1}, we observe that non-atomic contracts are characterized by an average number of swap events close to one and a relatively small set of interacting addresses. This observation is largely consistent with our initial assumptions during feature design.}
    \label{fig:decision_tree}
\end{figure*}

Figure~\ref{fig:decision_tree} shows a representative decision tree from the random forest.
The value array at each node in the tree contains two coordinates: the first represents the number of non-atomic MEV samples, and the second represents the number of other samples.  
We observe that the average number of swap events per transaction is the most indicative feature for detecting non-atomic contracts, with most non-atomic contracts exhibiting close to one swap event per transaction.
\cleardoublepage

\end{document}